\documentclass[11pt,showpacs]{revtex4}
\def\comment#1{}
\usepackage{amsmath}
\usepackage{amssymb}
\usepackage{graphicx}
\usepackage{epsfig}
\def\beq{\begin{equation}}
\def\eeq{\end{equation}}
\def\bea{\begin{eqnarray}}
\def\eea{\end{eqnarray}}

\usepackage{color}

\begin{document}

\title{Neutrinos decoupled from $\beta$-processes and supernova explosion}

\author{R. Mohammadi, Remo Ruffini and She-Sheng Xue}
\affiliation{ICRANet, P.zza della Repubblica 10, I--65122
Pescara, Physics Department and ICRA, University
of Rome, {\it La Sapienza} %P.le Aldo Moro 5, I--00185
Rome, Italy}
%\affiliation{$^{(b)}$Department of Physics, Isfahan University of Technology, Isfahan 84156-83111, Iran}

\date{Received version \today}

\begin{abstract}

Based on the gravitational collapse time-scale is larger than the weak interaction time-scale at core densities $\rho > 10^{11} {\rm gr}/ {\rm cm}^{3}$, we approximately use the $\beta$-equilibrium condition and particle number conservations to calculate the number and energy densities of neutrino sphere in the process of gravitational core collapse towards the formation of a proto-neutron star.
We find that at core densities $\rho_{\rm dec} > 10^{12} {\rm gr}/ {\rm cm}^{3}$,
the $\beta$-equilibrium condition cannot be satisfied consistently with charge, baryon and lepton
number conservations, leading to the presence of excess neutrinos decoupling from the $\beta$-equilibrium. These excess neutrinos interact with nucleons and electrons via the neutral current channel only and their diffusion time is about $10^{-2}\,$ sec. The excess neutrino flux could play an important role in an Supernova explosion, provided the fraction of excess neutrinos over all neutrinos is at least one present.
\end{abstract}

\maketitle
%%%%%%%%%%%%%%%%%%%%%%%%%%%%%%%%%%%%%%%%%%%%%%%%%%%%%%%%%%%%%%%
%%%%%%%%%%%%%%%%%%%%%%%%%%%%%%%%%%%%%%%%%%%%%%%%%%%%%%%%%%%%%%%
%\vskip0.1cm
\noindent
{\bf Introduction.}
\hskip0.1cm

A great effort has been made to understand the phenomenon of a core-collapse supernova for a long time.
It is known that the dominant weak interaction process altering the composition of the core matter in during collapse is $\beta$-process (electron capture and neutrons decay) on free and bound nucleons, which proceeds at a rate sufficient to produce a large number of neutrinos. In the core at densities larger than $10^{11} {\rm gr}/ {\rm cm}^{3}$, these neutrinos are trapped and thermalized, leading to electron capture equilibrium over time
scales shorter than the characteristic dynamical time scales for collapse \cite{beth, murphy,ar77}. The collapse continues essentially homologously \cite{rev7}, until nuclear densities reach
$10^{14}{\rm gr}/{\rm cm}^3$. Since nuclear matter has a much lower compressibility, the homologous core decelerates
and bounces in response to the increased nuclear matter pressure. This drives a shock wave into the
outer core, i.e. the region of the iron core which lies outside of the homologous core and in the meantime
has continued to fall inwards at supersonic speed \cite{rev,beth}.
If the shock wave were to propagate outward without stalling and make an explosion with energy about $\sim10^{51}$ergs,
it would be a successful {\it prompt hydrodynamical explosion}, but all of the realistic models completed to date suggest that this mechanism
does not occur at least for massive collapsing iron core, because the shock wave loses energy in dissociating iron nuclei when this shock passes through outside matter (the outside core matter includes iron group nuclei). Then the shock wave is enervated and loses its energy in the form of electron neutrinos, finally the shock wave would halt its outward motion \cite{beth,rews1,janka2007}.

Wilson proposed the {\it delayed mechanism} \cite{wilson85} that neutrino flux has a time scale much longer than {\it
prompt hydrodynamical explosion} and revives the stalled shock wave by the charged current absorption of electron neutrinos and anti-neutrinos.
However, in this mechanism, at least $99\%$ of the binding energy of the neutron star ($\sim10^{53}$ergs) comes out in neutrinos, which is $100$ times that needed for the shock wave to give a powerful supernova explosion \cite{rev,rews1}. In addition, compared with the time scale of the shock wave, the diffusion time of these neutrinos is too much longer to revive the shock wave at proper time \cite{rev}.
The main problem of delayed mechanism is then channeling some small fraction of the neutrino energy to the proper place and at the proper time to cause the explosion, several attempts have been made to find a solution to the problem \cite{rews1,janka2007,janka}.
In this letter, we study a possible solution to the problem.

 %Most neutrinos are engaged in the $\beta$-processes and their diffusion time is about a few seconds, cooling down a new born proto-neutron star.
%\vskip0.1cm
\noindent
{\bf Weak interaction and adiabatic gravitational collapse.}
\hskip0.1cm
In the standard model of particle physics, neutrinos interact with electrons and nucleons via charged and neutral current processes,
\begin{eqnarray}
    e^{-}+p &\rightleftharpoons & n+\nu_{e},\label{beta}\\
    \nu+(e,\,N) &\rightarrow& \nu+(e,\,N).\label{neut_cur_l}
\end{eqnarray}
The cross-section of the dominate $\beta$-processes (\ref{beta})
$\sigma_{_{\rm CC}}\approx 9.75\times10^{-42}(\bar E_\nu/10\,{\rm MeV})^2{\rm cm}^{2}
$
where $\bar E_\nu$ is the neutrino mean-energy. Neutrinos are left-handed, interacting only with left-handed quarks ($u,d$) inside nucleons and electrons by exchanging  charged gauge bosons $W^\pm$.
The cross-section of the neutrino-electron channel (\ref{neut_cur_l}), $\sigma_{_{\rm NC}}^{e\nu}\approx 0.01 \sigma_{_{\rm CC}}$ for $\bar E_\nu\sim 10$ MeV \cite{thompson2004,chang-book}.
The cross-section of the neutrino-nucleon channel (\ref{neut_cur_l}) is
\begin{equation}\label{cross-Nn}
 \sigma_{_{\rm NC}}\approx\sigma_{_{\rm CC}}\left[(g^u_L)^2+(g^d_L)^2+\frac{1}{3}\big((g^u_R)^2+(g^d_R)^2\big)\right],
\end{equation}
where $g^u_L=1/2-(2/3)\sin^2\theta_W$, $g^d_L=-1/2+(1/3)\sin^2\theta_W$, $g^u_R=-(2/3)\sin^2\theta_W$ and $g^d_R=(1/3)\sin^2\theta_W$
are left-handed neutrino gauge couplings to left- and right-handed quarks ($u,d$) by exchanging  neutral gauge boson $Z^\circ$ ($\sin^2\theta_W\simeq 0.25$) \cite{chang-book}.
$\sigma_{_{\rm NC}}\approx \sigma_{_{\rm CC}}/4$,
the $\beta$-processes (\ref{beta}) have a larger probability than neutral current processes (\ref{neut_cur_l}).

\begin{figure}[!t]
  \includegraphics[width=0.6\columnwidth]{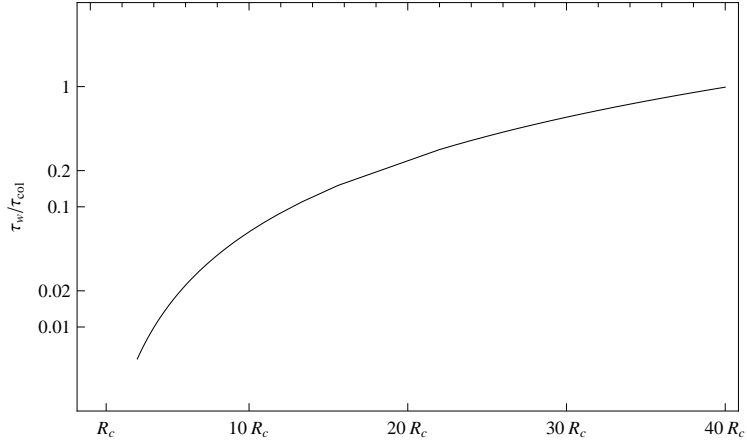}
  \caption{In the range $R_{\rm dec}< R < R^*$, the ratio $t_{\rm weak}/t_{\rm coll}$ is plotted  as a function of the collapsing radius $R/R_c$.}\label{adiabatic}
\end{figure}

We consider a collapsing stellar core of radius $R$ and mass $M\simeq M_\odot$, the collapsing time-scale $t_{\rm coll}=(\dot R/R)^{-1}\approx R/c$ is much larger than the weak-interaction time scale $
t_{\rm weak}\approx  (c\,\sigma_{_{\rm CC}} n_n)^{-1}$ ($t_{\rm coll}\gg t_{\rm weak}$), at nucleon densities $\rho \gtrsim 10^{11} {\rm gr}/ {\rm cm}^{3}$ for $R \lesssim R^*\approx20 R_c$ and $\bar E_\nu\sim 10\,$MeV (see Fig.~\ref{adiabatic}). This implies that gravitational collapse
could be approximately treated as slowing varying adiabatic process with respect to the $\beta$-processes (\ref{beta}), which is in agreement with other studies (see for review \cite{rev0}). In this {\it adiabatic approximation}, at each collapsing radius $R$, the $\beta$-equilibrium is assumed to be {\it locally and instantaneously} established in the characteristic space-time variations $\Delta R$ and $\Delta t$, determined by the equation of gravitational collapse. The analogous discussion can be applied for other macroscopic processes. Based on this {\it adiabatic approximation}, we study neutrino emission in gravitational collapse.
%\vskip0.1cm

\noindent
{\bf $\beta$-equilibrium and neutrino sphere.}
\hskip0.1cm
Suppose that the stellar core is composed by complete degenerate gases of electrons, neutrinos, protons and neutrons for their Fermi momenta being much larger than temperature.
The $\beta$-equilibrium condition is
\begin{eqnarray}
\mu_{n} + \mu_\nu = \mu_{p} + \mu_e,
\label{Equibeta}
\end{eqnarray}
where chemical potentials
\begin{eqnarray}
\mu_{n,p,e,\nu} &= &\sqrt{(cp^F_{n,p,e,\nu})^2+m_{n,p,e,\nu}^2c^4}.
\label{chemnv}
\end{eqnarray}
Suppose that particles homogeneously distribute within the core of volume $V=4\pi R^3/3$,  their densities $n_{n,p,e,\nu}=N_{n,p,e,\nu}/V$ and Fermi momenta $p^F_{n,p,e,\nu}=(3\pi^2)^{1/3}\hbar\, n_{n,p,e,\nu}^{_{1/3}}$. $A=N_n+N_p$ and $L=N_e+N_\nu$ are the total conserved baryon and lepton numbers, local neutrality requires $n_p=n_e$. The variation of neutrino number is related to nucleon one,
$\Delta N_\nu=\Delta N_n=-\Delta N_p$. These equations completely determine the ratio $A/N_p$ and $n_{n,p,e,\nu}(R)$ as functions of the core radius $R$.
During collapse, the variation of gravitational binding energy $\Delta \mathcal E_g  \approx - 3(GM^2/R)(\Delta R/R)$ \cite{Weinberg}, and
it is believed that about $99\%$ of this energy is converted to the neutrino energy \cite{arnett}.
Neutrinos are trapped inside the core for large opacity
\begin{eqnarray}
  \tau_\nu \approx  (\sigma^{_{\rm CC}} +\sigma^{_{\rm NC}})n_n R\approx \sigma^{_{\rm CC}} n_n R \gtrsim 1,
  \label{t-p1}
\end{eqnarray}
for $R\lesssim R^*\approx20 R_c$, coinciding with the valid range of {\it adiabatic approximation}.
\comment{
and about $1\%$  of this energy is released to the energy of neutrinos produced by
$\beta$-processes (\ref{beta}) during collapse, which is adopted for different models considering collapsing dynamics
and observational data (see \cite{arnett,rews,rews1,arrnet-sch} and \cite{vassini}).
}

\begin{figure}[!t]
  \includegraphics[width=0.6\columnwidth]{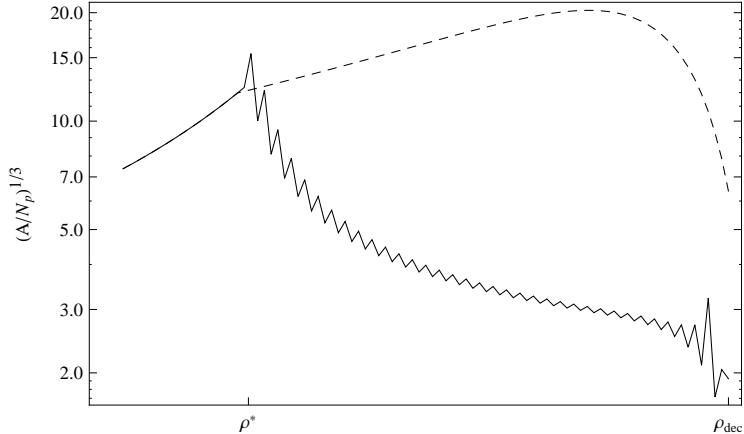}
  \caption{In the range $\rho^*<\rho<\rho_{\rm dec}$, the ratio $(A/N_p)^{1/3}$ is plotted (solid line) as a function of the core density $\rho$.  The dashed line is for the case all neutrinos escape $\mu_\nu=0$.  }\label{apr_trapped}
\end{figure}

\begin{figure}%[!h]
\begin{center}
\includegraphics[width=3.1in]{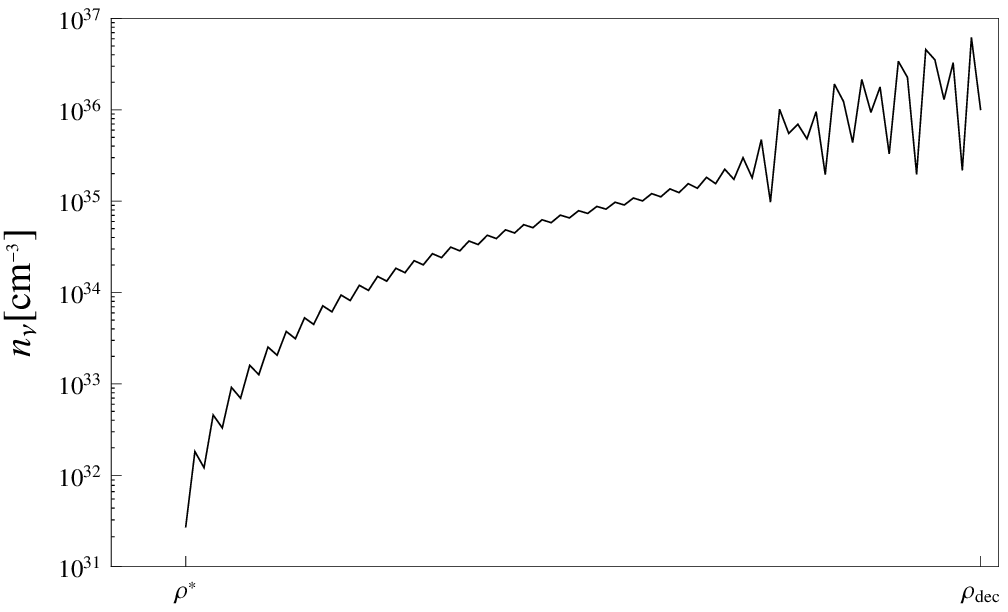}\includegraphics[width=3.1in]{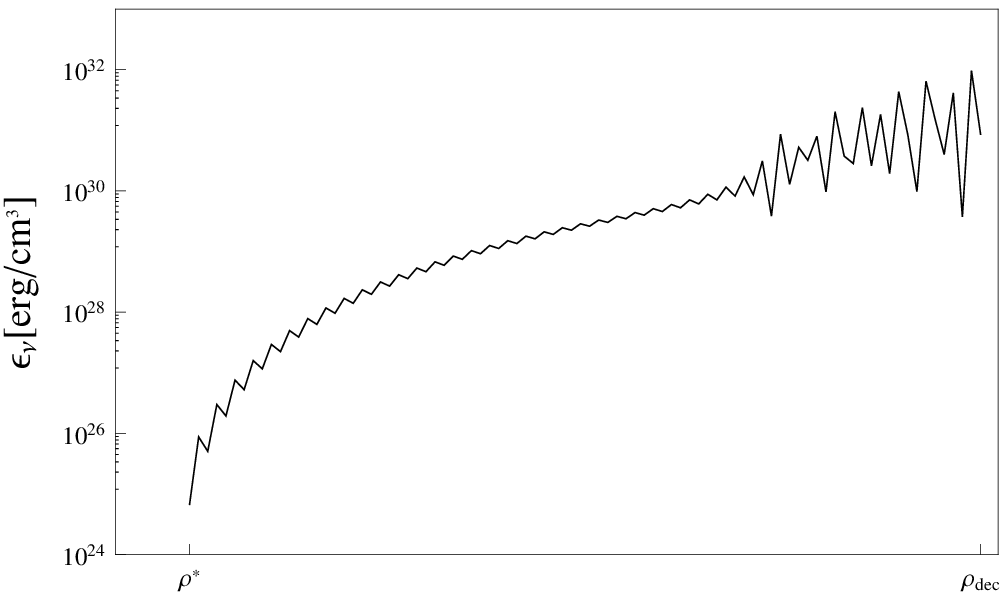}
\caption{In the range $\rho^*<\rho<\rho_{\rm dec}$ and $\rho_{\rm dec}\approx \rho_{\rm nucl}(R_c/R_{\rm dec})^3$ where $R_{\rm dec}\simeq(2\sim5)R_c$, we plot the homogeneous number-density $n_{\nu}$ (left panel) energy-density $\epsilon_\nu$ (right panel) of neutrino sphere formed at the  core density $\rho$.}
\label{number-d-t}
\end{center}
\end{figure}

Using Eqs.~(\ref{Equibeta}-\ref{t-p1}), we determine a neutrino sphere (see Figs.~\ref{apr_trapped} and \ref{number-d-t}) by numerically calculating the ratio $A/N_p$, particle number-density $n_{n,p,e,\nu}$ and energy-densities $\epsilon_{n,p,e,\nu}$ as functions of the core radius (or core density), starting from the radius $R^*\approx 20 R_c$ ($\rho^*\approx 5\times 10^{10}{\rm gr}/{\rm cm}^{3}$) to the radius $R_{\rm dec}\approx (2\sim 5)R_c$ ($\rho_{\rm dec}\approx \rho_{\rm nucl}(R_c/R_{\rm dec})^3$), which will be clarified below. The obtained value $A/N_p$ at the radius $R_{\rm dec}\sim R_c$ (a new
born proto-neutron star radius) is consistent with numerical simulation
results $Y_{e,p}=N_{e,p}/A\approx 0.3-0.5$ \cite{beth,ar77,rews1,palbook} in agreement with
observation data \cite{flux}.
It is shown that at the radius $R_{\rm dec}$, $\tau_\nu \sim 10^{2-3}$, $n_\nu\sim 10^{36-37}/{\rm cm}^3$, $\epsilon_\nu\sim 10^{31-32}{\rm ergs}/{\rm cm}^3$, and the total neutrino energy and number are $10^{52-53}$egrs and $10^{56-57}$. The neutrino mean-energy $\bar E_\nu= \epsilon_{\nu}/n_{\nu}\approx 10$MeV.

%%%%%%%%%%%%%%%%%%%%%%%%%%%%%%%%%%%%%%%%%%%

%\vskip0.1cm
\noindent
{\bf Excess neutrinos over the $\beta$-equilibrium.}
\hskip0.1cm
The numerical calculation shows that at the radius $R_{\rm dec}\approx (2\sim 5)R_c$, where the neutrino chemical potential is so large that non solution fully fills the $\beta$-equilibrium condition (\ref{Equibeta}) with baryon, lepton and charge conservations.  In fact, it is a necessary condition that the neutrino chemical potential should be smaller than electron one, $\mu_\nu<\mu_e$, because the neutron chemical potential is larger than proton one $\mu_n >\mu_p$. In another word, $N_{\nu}<N_e/2$, where the factor $1/2$ is due to different neutrino and electron spin degeneracies.
We might introduce the radius $R_{\rm dec}$ determined by the critical condition $\mu_\nu(R_{\rm dec})= \mu_e (R_{\rm dec})$, then Eq.~(\ref{Equibeta}) yields the critical condition
\begin{equation}
N_n \simeq 2N_{\nu}=N_e=N_p=A/2,
\label{cri_nu}	
\end{equation}
at the radius $R_{\rm dec}$. We analytically calculate
the neutrino chemical potential $\mu_\nu(R_{\rm dec}) = \mu_e (R_{\rm dec})\approx250\, {\rm MeV}(R_c/R_{\rm dec})$, number-density $n_{\nu}(R_{\rm dec}) \simeq  2\times 10^{38}(R_c/R_{\rm dec})^3 {\rm cm^{-3}}$ and energy-density $\epsilon_{\nu}(R_{\rm dec}) \simeq  2\times 10^{34}(R_c/R_{\rm dec})^4 {\rm erg~cm^{-3}}$, consistently with the numerical result at the radius $R_{\rm dec}$ (see Figs.~\ref{apr_trapped} and \ref{number-d-t}).

When the collapsing radius $R< R_{\rm dec}$, $N_{\nu} > N_{n}/2$ then neutrons cannot absorb all neutrinos via the $\beta$-process $n+\nu\rightarrow p+e$. As a consequence, excess neutrinos $N^{^{\rm EX}}_{\nu}$ over the $\beta$-equilibrium, namely those neutrinos decoupled from the $\beta$-processes (\ref{beta}), must be present in the neutrino sphere. This can be understood from the microscopic point of view, the difference between the
neutrino emissivity and the absorption by the thermal system of electrons, protons and neutrons
via the $\beta$-processes (\ref{beta}) is given by \cite{thompson2004}
\begin{eqnarray}
  \mathcal{C}_{\beta}= \int d\omega_\nu\,\frac{\kappa_\nu/\omega_\nu}{1-{\mathcal F}_\nu^{'}}\,(B_\nu- I_\nu),\quad \mathcal{Q}_{\beta}= \int d\omega_\nu\,\frac{\kappa_\nu}{1-{\mathcal F}_\nu^{'}}\,(B_\nu- I_\nu),
  \label{net_coll0}
\end{eqnarray}
where $\kappa_\nu$, $I_\nu$ and $B_\nu$ are the absorptive opacity, the specific intensity and the black-body function for neutrinos. In Eq.~(\ref{net_coll0}),
$
   \mathcal{F}_\nu^{'}\equiv  [e^{[\omega_\nu-(\mu_e+\mu_p-\mu_n)]/kT}+1]^{-1}
$ and $\omega_\nu\approx |\mathbf{p}_{\nu}|$.
When the detailed balance is established, i.e., the neutrino absorption and emission rates are exactly equal, neutrinos are in thermal equilibrium with the system via the $\beta$-processes, the $\beta$-condition (\ref{beta}) is fully satisfied and $\mathcal{C}_{\beta}$ ($\mathcal{Q}_{\beta}$) vanishes, leading to the neutrino black body distribution
$
   \mathcal{F}_\nu^{'} = \mathcal{F}_\nu= [e^{(\omega_\nu-\mu_\nu)/kT}+1]^{-1}
$
and $B_\nu = I_\nu$.
This is the case for $R > R_{\rm dec}$, we use $\beta$-equilibrium condition (\ref{beta}) to calculate number- and energy-densities of the neutrino sphere.
In the case $R\lesssim R_{\rm dec}$, the $\beta$-equilibrium condition (\ref{beta}) cannot be satisfied and
$
   \mathcal{F}_\nu\not = \mathcal{F}_\nu^{'},
$
and the neutrino chemical potential $\mu^{\rm non}_\nu$ is different from the $\beta$-equilibrium one
$\mu_\nu$ ($\mu^{\rm non}_\nu >\mu_\nu$).
As a result, $\mathcal{C}_{\beta},\mathcal{Q}_{\beta}\not= 0$ indicates neutrinos decoupled from the $\beta$-processes (\ref{beta}), namely the excess of neutrinos over the $\beta$-equilibrium. We call the radius $R_{\rm dec}\approx  (2 \sim 5)R_c$
decoupling radius, corresponding
core densities $\rho_{\rm dec} \approx (10{\rm Km}/R_{_{\rm dec}})^3\rho_{\rm nucl}$, where $\rho_{\rm nucl}$ is the nuclear density and  $R_{_{\rm dec}}= (20 \sim 50)$Km.

\begin{figure}
  \includegraphics[width= 4in]{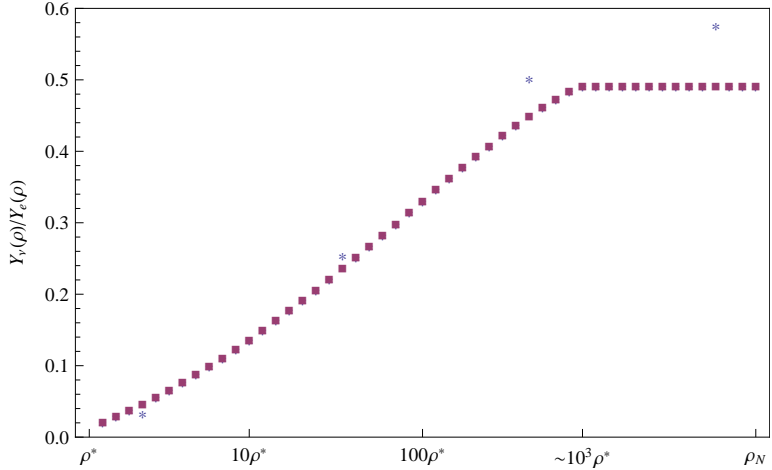}\\
  \caption{$Y_\nu(\rho)/Y_e(\rho)$ is plotted as function of core density ($\ast$ and quadrangle points are results of \cite{G15} and fitting formula \cite{lie2005}), $\rho^*\sim 5\times10^{10}{\rm gr/cm^3}$ is the density which is started the trapping of neutrinos.}\label{n-f-1}
\end{figure}

We would like to see whether this critical condition (\ref{cri_nu}) appears in calculations by numerical simulations.  Using numerical simulation data \cite{G15} and fitting function \cite{lie2005} for electron fraction $Y_e$, in Fig.~(\ref{n-f-1}), we plot the ratio $Y_\nu/Y_e$ (neutrino fraction over electron fraction) in terms of core density $\rho$. We find at core densities $\sim  10^3$ this ratio $Y_\nu/Y_e \rightarrow 0.5$, consistently with the decoupling density $\rho_{\rm dec}$ obtained in Fig.~\ref{apr_trapped} and Eq.~(\ref{cri_nu}).

%\vskip0.1cm
\noindent
{\bf Excess neutrino luminosity and diffusion time.}
\hskip0.1cm
These excess neutrinos decouples from the $\beta$-processes, then the cross-section of excess neutrinos via the $\beta$-processes vanishes, $\sigma^{^{\rm EX}}_{_{\rm CC}}=0$.
This implies that all left-handed quark fields inside nucleons are fully engaged with left-handed neutrinos via the $\beta$-processes when $R\lesssim R_{\rm dec}$. Therefore, excess neutrinos scatter only with right-handed quarks via neutral current interaction, i.e., the $g_R^{u,d}$-terms in Eq.~(\ref{cross-Nn}), yielding the cross-section $\sigma^{\rm R}_{_{\rm NC}}\approx 0.01 \sigma_{_{\rm CC}}$.
As a result, the excess neutrino opacity is only due to the scattering via right-handed neutral current channels (\ref{neut_cur_l}),
\begin{eqnarray}
\tau^{^{\rm EX}}_\nu=\sigma^{^{\rm EX}}_{_{\rm NC}}n^{{\rm R}}_{_{\rm N}} R\approx 0.01\, \tau_\nu,\quad \sigma^{^{\rm EX}}_{_{\rm NC}}=\sigma^{e\nu}_{_{\rm NC}}+\sigma^{{\rm R}}_{_{\rm NC}}\label{neut_opa}
\end{eqnarray}
where $n^{{\rm R}}_{_{\rm N}}\approx n_{_{\rm N}}/2$ the nucleon number-density, which accounts only right-handed quark fields inside nucleons.
The diffusion time $t^{^{\rm EX}}_{\rm diff}$ can be estimated by (see for example \cite{rev0}),
\begin{equation}\label{diff}
t^{^{\rm EX}}_{\rm diff}\simeq  3\frac{\lambda^{^{\rm EX}}_{\nu}}{c} N_{\rm coll}\sim 3\frac{R_{\rm dec}}{c}\left(\frac{R_{\rm dec}}{\lambda^{^{\rm EX}}_{\nu}}\right)
=3\frac{R_{\rm dec}}{c}\tau^{^{\rm EX}}_{\nu},
\end{equation}
where the collision number $N_{\rm coll}\sim (R_{\rm dec}/\lambda^{^{\rm EX}}_{\nu})^2$, and  the mean-free path  of excess neutrinos $\lambda^{^{\rm EX}}_{\nu}=1/(\sigma^{^{\rm EX}}_{_{\rm NC}}n^{{\rm R}}_{_{\rm N}})$.
At the radius $R_{\rm dec}$, we obtain the excess neutrino diffusion time $t^{^{\rm EX}}_{\rm diff}\sim 0.01\,$sec and averaged diffusion velocity $v^{^{\rm EX}}_{\rm diff}=R_{\rm dec}/t^{^{\rm EX}}_{\rm diff}\sim 10^{-1}$c.
The outgoing energy-flux of these excess neutrinos  $F_{\nu}= \mathcal {Q}_\beta = \eta_{_{\rm EX}} \,\epsilon_{\nu}\, v^{^{\rm EX}}_{\rm diff}$, where the parameter $\eta_{_{\rm EX}} (\eta_{_{\rm EX}}<1)$ presents the fraction of excess neutrinos over the $\beta$-equilibrium. The corresponding luminosity is
\begin{eqnarray}\label{l1}
    L^{^{\rm EX}}_{\nu}(R) &=& 4\pi  R^2  \eta_{_{\rm EX}}\,\epsilon_{\nu}\,v^{^{\rm EX}}_{\rm diff}\nonumber\\
    &=& 5\times10^{57}\eta_{_{\rm EX}}\,\left(\frac{\lambda^{^{\rm EX}}_\nu}{R}\right)\left(\frac{R_c}{R}\right)^2{\rm erg/sec},
\end{eqnarray}
and $L_{\nu}(R_{\rm dec})\approx  5\times 10^{54}\eta_{_{\rm EX}}\,{\rm erg/sec}$, increasing as the collapsing radius $R$ decreases in the range $R_c\lesssim R\lesssim R_{\rm dec}$.

We try to estimate the fraction $\eta_{_{\rm EX}}$ of excess neutrinos produced in the range $R_c\lesssim R\lesssim R_{\rm dec}$.
Due to their short diffusion time $\sim 10^{-2}$ sec, excess neutrinos are supposed to completely emit away when the collapsing radius $R$ varies from $R_{\rm dec}$ to the radius $R_c$ of a
new born proto-neutron star, in which protons, neutrons, electrons and $\beta$-neutrinos are in the $\beta$-equilibrium at the critical condition $ N_n\approx N_p=N_e=2 N_\nu$. Under this assumption,
we calculate the energy variations $\Delta E_{p,n,e,\nu}^{\rm int}=E_{p,n,e,\nu}^{\rm int}(R_c)-E_{p,n,e,\nu}^{\rm int}(R_{\rm dec})\simeq(0.9\sim 1.5)\times 10^{53}{\rm ergs}$,
as well as the energy emission of excess neutrinos (\ref{l1}).
The variation of gravitational binding energy $\Delta \mathcal E_g  \approx - 3(GM^2/R)(1-R_c/R_{\rm dec})\approx -(1\sim 1.7)\times 10^{53}{\rm ergs}$.
We obtain $\eta_{_{\rm EX}} < 0.1$ by the total energy conservation.

%\vskip0.1cm
\noindent
{\bf Excess neutrinos and Supernova explosion.}
\hskip0.1cm
The necessary conditions for an Supernova explosion due to neutrino flux are:
(i) a steady neutrino flux occurs in the range between a proto-neutron star (PNS) radius $R_c$ (shock wave starts) and shock radius $R_{\rm shock}\sim 200$km (shock wave stalls), corresponding to the shock time-interval $t_{\rm sh}\sim 0.1$ sec; (iii) the steady neutrino luminosity exceeds a threshold luminosity $L^{^{\rm crit}}_{\nu}$. The theoretical and numerical studies show in order to have a powerful explosion, it needs the neutrino luminosity threshold $L^{^{\rm crit}}_{\nu}\approx 5\times 10^{52}(\dot{M}/\dot{M}_\odot)\,$erg/sec to overcome the gravitational pressure of infilling matter. The rate of infilling matter $\dot{M}/\dot{M}_\odot\sim 0.1$
decreases during the shock time interval, and increases after shock wave stalls \cite{janka2001}.
If the neutrino diffusion time is smaller than or the same order as the shock wave time, the outgoing neutrino flux is in the same direction of outgoing shock wave with small rate $\dot{M}/\dot{M}_\odot$. This implies that (i) the critical luminosity
$L^{^{\rm crit}}_{\nu}$ becomes small; (ii) neutrinos have much more probability to interact with infilling matter and transfer their energy-momenta to make a powerful explosion \cite{janka,janka2001,janka2007}.

We are in position of marking the following remarks on the excess neutrino luminosity (\ref{l1}) for an supernova explosion:
\begin{enumerate}
\item The excess neutrino luminosity turns on at the radius $R_{\rm dec}\sim (2-5) R_c$, indicating that an instability is triggered. The shock wave occurs around this radius;

\item The time scale of excess neutrino luminosity is about $10^{-2}\,$sec., the same order of shock wave time scale $t_{\rm sh}$, indicating that the excess neutrino luminosity starts to act for explosion almost at the same time as the shock wave starts, rather than after shock wave stalls;

\item If the fraction of excess neutrinos overall neutrinos is about one percent ($\eta_{_{\rm EX}}\sim 1\%$), the excess neutrino luminosity (\ref{l1}) is larger than the threshold luminosity $L^{^{\rm crit}}_{\nu}\sim 5\times 10^{52}$erg/sec, satisfying the necessary condition for an supernova explosion.
\end{enumerate}
These properties of excess neutrino luminosity might give a solution to
the main problem of delayed mechanism: channeling some small fraction ($\eta_{_{\rm EX}}\sim 10^{-2}$) of the neutrino energy to the proper place [$R_{\rm dec}\sim (2-5)\times 10^6\,$cm] and at the proper time ($t^{^{\rm EX}}_{\rm diff}\sim 10^{-2}\,$sec) to cause the supernova explosion. The energy of excess neutrinos can be estimated by $L_{\nu}(R_{\rm dec})\times t^{^{\rm EX}}_{\rm diff}\sim  \eta_{_{\rm EX}}10^{53}\,{\rm ergs}$. However, it should be mentioned that (i) the excess neutrino decoupling radius $R_{\rm dec}\sim (2-5) R_c$ is a low-limit, because there are other processes of neutrino productions which we do not consider; (ii) the excess neutrino diffusion time-scale $10^{-2}\,$sec is also a low-limit because it is obtained from the purely right-handed neutral current interaction; (iii) the excess neutrino fraction $\eta_{_{\rm EX}}< 0.1$ is an up-limit, because it is obtained by total energy conservation and we do consider all possible energy dispassive channels.
In the present model and calculations, we are not able to give the up-limit of the excess neutrino diffusion time-scale and the low-limit of  excess neutrino fraction $\eta_{_{\rm EX}}$. Thus, excess neutrinos might not play an important role in Supernova explosions for the following two cases: either the the excess neutrino diffusion time-scale is larger than the shock time-scale ($t_{\rm th}\sim 0.1\,$ sec) or the the excess neutrino fraction is so small that the excess neutrino luminosity is smaller than critical one $L^{^{\rm crit}}_{\nu}$. Nevertheless, we will show that in cooperation with strong electric fields on stellar core surface \cite{rrx2011}, these excess neutrinos can at least play an important role for enhancing electric field \cite{hrx2011} and trigging electron-positron pair productions \cite{mrx2012}.

%\vskip0.1cm
\noindent
{\bf Neutrino emission in cooling phase.}
\hskip0.1cm
In the neutrino sphere, in addition to excess neutrinos, most neutrinos participate the $\beta$-equilibrium (\ref{beta}) and we call them $\beta$-neutrinos to distinguish them from excess neutrinos.
Their diffusion time can be estimated by
\begin{equation}\label{diff1}
t^{^{\rm T}}_{\rm diff}\simeq 3\frac{R_{\rm dec}}{c}\tau^{^{\rm T}}_{\nu},~~~~ \tau^{^{\rm T}}_{\nu}=n_{_{\rm N}}(\sigma_{_{\rm CC}}+\sigma_{_{\rm NC}})\approx n_{_{\rm N}}\sigma_{_{\rm CC}}.
\end{equation}
which is about a few seconds ($t^{^{\rm T}}_{\rm diff}\gtrsim 1\,$sec) for the core radius $R_c\sim 10^6$cm and opacity $\tau_\nu\sim 10^3$. The diffusion time (\ref{diff1}) of $\beta$-neutrino flux is much larger than the shock time $t_{\rm sh}$, the outgoing $\beta$-neutrino flux starts after the shock wave stalls and the infilling mater rate $\dot{M}/\dot{M}_\odot$ increases. This implies that (i) the threshold luminosity $L^{^{\rm crit}}_{\nu}$ becomes large; (ii) $\beta$-neutrinos have not much probability to interact with infilling matter and transfer their energy-momenta to make a powerful explosion. Therefore $\beta$-neutrinos cannot be relevant for the delayed mechanism to revive the shock wave.

Because of their short diffusion time ($10^{-2}\,$sec), excess neutrinos are assumed to completely emit when a new born proto-neutron star is formed at the radius $R_c$ after an Supernova explosion. The number of major $\beta$-neutrinos left over inside the neutrino sphere is about $A/2$, which is approximately given by the critical condition (\ref{cri_nu}).
Due to the radial gradient of pressure and particle density inside the core, these $\beta$-neutrinos diffuse outwards and emit from the core, leading to the cooling of new born proto-neutron stars. In this cooling phase, while $\beta$-neutrinos are diffusing out of the core, the neutrino production via the process of electron capture (\ref{beta}) takes place. As a consequence, the value $A/N_p$ increases from the value $A/N_p\approx 2$ for neutrinos trapping to the value $A/N_p\approx 250$ for all neutrinos escaping at the end of cooling phase (see Fig.~\ref{apr_trapped}). Thus, the total number of neutrinos produced by the process of electron capture should be about $A/2$. The total number of emitted neutrinos in the cooling phase should be about $A$, which is the sum over $\beta$-neutrinos and neutrinos produced by electron capture process from $A/N_p\sim 2$ to $A/N_p\sim 250$.
Carrying major neutrino energy and number in the entire neutrino sphere, $\beta$-neutrinos play an important role in the cooling phase of a new born proto-neutron star \cite{arnett,rews1}.
As mentioned, excess neutrino energy is about $\eta_{_{\rm EX}}10^{53}\,{\rm ergs}$. Then $\beta$-neutrino energy should be about $(1-\eta_{_{\rm EX}})10^{53}\,{\rm ergs}$. This is the predominant cooling mechanism immediately after formation proto-neutron star with a timescale of seconds, and the total energy carried away by $\beta$-neutrinos and neutrinos produced by electron capture process should be the major part (about $99\%$) of the gravitational binding energy \cite{arnett,rews1}.

%\vskip0.1cm
\noindent
{\bf Some remarks.}
\hskip0.1cm
In this Letter, we only consider the core of collapsing star at radii $R < R^*\approx 20 R_c$ ($\rho^*\sim 10^{11}{\rm gr}/{\rm cm}^3$) on the basis of homologous collapse \cite{rev7}. However, we assume that the core density $\rho$ is spatially homogeneous and its sharp boundary described by a Heaviside step function, i.e.,  $\rho = \delta(r-R)M/V$, neglecting the variation of core density at the boundary. This approximation leads to the effective mass inside the core is larger than realistic one, then energy- and number-densities of neutrino sphere calculated are probably larger than that in realistic case. In addition, ignoring nuclei of iron group in the core, we adopt a model of completely degenerate free electrons, protons and neutrons to calculate their chemical potentials. This approximation affects on the calculations of the ratio $A/N_p$ and neutrino productions. Under these approximations, we show the possible presence of excess neutrinos decoupled from the $\beta$-processes, and obtain the preliminary results of the decoupling radius $R_{\rm dec}$, excess neutrino fraction $\eta_{_{\rm EX}}$, excess neutrino diffusion time $t^{^{\rm EX}}_{\rm diff}$ and excess neutrino luminosity $L^{^{\rm EX}}_\nu$, which could be a possible solution to the main problems of delay mechanism for supernova explosion. We also briefly discuss $\beta$-neutrinos and neutrinos produced by electron capture process in the cooling phase of new born proto-neutron stars. The detailed calculations to obtain these preliminary results of this Letter will be presented in a lengthy article \cite{mrx2012}.
Needless to say, these preliminary results are necessary to be verified by other approaches, in particular numerical approaches.

%\vskip0.1cm
\noindent
{\bf Acknowledgment.}
\hskip0.1cm
R.~Ruffini and S.-S.~Xue thank D.~Arnet and A.~Mezzacappa for many discussions in Les Houches. R.~Mohammodi is grateful to the hospitality extended to him while he visits the ICRANet in Pescara.

%%%%%%%%%%%%%%%%%%%%%%%%%%%%%%%%%%%%%%%%%%%%%%%%%%%%%%%%%%%%%%%%%%%%%%%%%%%%%%%%%%%%%%%%%%%%%%%%%%%%%%%%%%%%%%%%%%%%%%%%%

%{99}
\end{document}